%% file: perf_ano_sol.tex
\documentclass[a4paper]{article}
\usepackage{RR}
\usepackage{hyperref}
\usepackage{amsfonts}
\usepackage{amsmath}
\usepackage{graphicx}
\usepackage{algorithm2e}
\usepackage{multirow}
\RRdate{July 2006}
\RRauthor{
Tahiry RAZAFINDRALAMBO\thanks{tahiry.razafindralambo@insa-lyon.fr}%
  \and
Isabelle GU\'ERIN-LASSOUS\thanks{isabelle.guerin-lassous@insa-lyon.fr}%
\and
Luigi IANNONE\thanks{luigi.iannone@lip6.fr}
\and
Serge FDIDA\thanks{serge.fdida@lip6.fr}
}
\authorhead{Razafindralambo, Gu\'erin-lassous, Iannone, Fdida }
\RRtitle{Agr\'egation Dynamique de Paquets pour R\'esoudre \\ l'Anomalie de Performance des R\'eseaux sans Fils 802.11 }
\RRetitle{Dynamic Packet Aggregation to Solve \\ Performance Anomaly in 802.11 Wireless Networks}
\titlehead{PAS: Performance Anomaly Solution}
\RRnote{}
\RRnote{}
\RRresume{L'anomalie de performance est un probl\`eme bien connu du standard 802.11. Il est aussi l'un des plus \'etudi\'es. Ces derni\`eres ann\'ees des solutions permettant de r\'esoudre ce problème, telles que la fragmentation de paquet, l'adaptation de l'algorithme de backoff, ou l'agr\'egation d'envois de paquets durant un temps donn\'e, ont \'et\'e developp\'ees. Dans ce papier nous proposons une solution au probl\`eme de l'anomalie de performance bas\'ee sur une agr\'egation des paquets en utilisant un intervalle de temps dynamique, qui d\'epend du temps d'occupation du m\'edium sans fil. Cette approche dynamique nous permet d'augmenter l'\'equit\'e, la r\'eactivit\'e, et d'\^etre dans certain cas plus efficace compar\'e aux autres solutions propos\'ees dans la litt\'erature.
}

\RRabstract{In the widely used 802.11 standard, the so called \emph{performance anomaly}  is a well known issue. Several works have tried to solve this problem by introducing  mechanisms such as packet fragmentation, backoff adaptation, or packet aggregation during a fixed time interval. 
In this paper, we propose a novel approach solving the performance anomaly problem by packet aggregation using a dynamic time interval, which depends on the busy time of the wireless medium. 
Our solution differs from other proposition in the literature because of this dynamic time interval, which allows  increasing fairness, reactivity, and in some cases efficiency.  In this article, we emphasize the performance evaluation of our proposal.}
\RRmotcle{R\'eseaux sans fil; IEEE~802.11; Anomalie de Performance.}
\RRkeyword{Wireless Networks; IEEE~802.11; Performance Anomaly.}
\RRprojets{ARES}
\RRtheme{\THCom} 
\URRhoneAlpes
\begin{document}
\makeRR   

\section{Introduction}
\input{intro.tex}

\section{The Performance Anomaly}
\label{sec:802}
\input{802overview.tex}

\section{Related work}
\label{sec:art}
\input{related.tex}

\section{PAS: a dynamic packet aggregation}
\label{sec:des}
\input{protocol.tex}

\section{A theoretical analysis}
\label{sec:theo}
\input{analytical.tex}

\section{Simulations results}
\label{sec:sim}
\input{simulation.tex}

\section{Conclusion}
\label{sec:conc}
\input{conclusion.tex}


\end{document}

%% file: intro.tex
Performance anomaly is a key issue in IEEE 802.11 multi-rate wireless
networks. 
It decreases the network global performance because of a bad time sharing
between stations transmitting at high bit rate (fast stations) and
stations transmitting at slow bit rate (slow stations).
This bad time sharing results in an unfair throughput, with slow
stations throttling fast stations' traffic \cite{Dud03}. 
Several solutions have been proposed in the literature to solve this 
problem. 
Some of them are based on a static and predefined time sharing 
between slow and fast stations by shaping the MTU 
(Maximum Transmission Unit) on a transmission rate basis. Other
approaches set the maximum amount of time a station can hold the
medium, like with the TXOP (transmit opportunity) introduced in the
IEEE 802.11e standard. Finally, other approaches try to adapt the
contention window size of IEEE~802.11, accordingly to the transmission
rate of the station. 

The main problem of existing solutions is that they are static
or centralized. In this paper, we tackle both issues, solving the performance
anomaly with a dynamic and distributed approach. 
Our solution is dynamic because it introduces a transmission time, 
similar to the TXOP, that changes depending on the perceived channel
occupancy, which in turns evolves with the traffic load of the network. 
Our solution is a distributed approach because each node computes locally
the maximal channel occupancy time, based on the active medium sensing 
provided by IEEE~802.11. 
Once a node gains access to the medium, it can send as many packets as 
allowed by the computed transmission time depending on the sensed 
maximal channel occupancy time. 

In this article, we emphasize the performance evaluation of our
approach. 
We propose an analytical evaluation of our protocol
in the classical scenario where all stations are within communication
range and a detailed simulation-based evaluation. We evaluate our
protocol in terms of efficiency and of fairness on many configurations
not limited to one-hop networks. We also compare our solution to
three different approaches that belong to the three main classes of solutions
solving the performance anomaly.   

The remaining of the paper is organized as follow. 
We give a short overview on the IEEE~802.11 access function and
describe the performance anomaly in Section~\ref{sec:802}.
In Section~\ref{sec:art} we propose a review of the existing 
modifications of the IEEE~802.11 that solve the performance anomaly. 
In Section \ref{sec:des} we describe our proposal. 
In Section \ref{sec:theo} we propose an analytical evaluation
for a specific topology while in Section~\ref{sec:sim} we describes the 
simulations carried out to evaluate the performances and the impact of
the different parameters of the proposed protocol on various
scenarios. 
Finally, we conclude the paper with the perspectives raised by this 
work in Section~\ref{sec:conc}.

%% file: 802overview.tex
The IEEE~802.11 standard~\cite{80211} provides a totally distributed 
medium access
protocol, called the Distributed Coordination Function (DCF). 
The DCF is part of the Carrier Sense Multiple Access with Collision Avoidance
(CSMA/CA) family. 
Emitters have to wait for the channel to become free before sending a
frame. 
When a frame is ready to be emitted, it is emitted after a fixed time 
interval called the {\em DIFS} (Distributed Inter Frame Space)
during which the medium shall stay idle. 
If the medium is or becomes busy during this interval, 
a random number called {\em backoff} out of an interval called
{\em Contention Window} ($CW$) is generated. 
This number indicates the time to be waited before transmitting. 
When the medium becomes idle again, the mobile waits for a DIFS 
before starting to decrement its backoff. 
When the medium becomes busy during the decrease, the process is stopped
and will be resumed later after a new DIFS with the remaining
backoff. 
As soon as the backoff reaches 0, the frame is emitted. 
Since collision detection is not possible, each unicast frame has to be
acknowledged. When a receiver successfully receives a frame, it waits for
a {\em SIFS} (Short Inter Frame Space) time and then emits the 
acknowledgment. 
The SIFS is shorter than the DIFS in order to give priority to acknowledgments
over data frames. 
The lack of the reception of an acknowledgment is considered as a 
collision. 
In that case, the $CW$ size is doubled and the same frame is re-emitted with
the same process described previously. 
If another collision happens, the $CW$ size is doubled again if it has
not yet reached the maximum value defined by the standard. 
After a fixed number of retransmissions, the frame is dropped and the 
$CW$ size is reset, as for a successful transmission.

Heusse \emph{et Al.}~\cite{Dud03} have shown that the presence of 
slow terminals in a multi-rate wireless network slows
down every other terminal. 
During the transmission of a slow terminal the medium is busy for a longer
period than during the transmission of a fast terminal.
Since 802.11 provides simple per-packet fairness in one-hop
networks, meaning that in a long period, each emitter statistically 
has sent the same number of frames. On a time basis, however, slow
terminals have occupied the channel for a longer period of time. 
This time unfairness that arise as soon as multiple rates are present, 
can lead to a loss of performance due to the existence of slow
transmissions.

%% file: related.tex
By letting both fast and slow stations to capture the channel 
for the same amount of time, the performance of IEEE~802.11 should be 
improved. 
The issue has been tackled in several different ways, with solutions 
placed at different levels of the protocols stack. 
Here we present the most relevant works that try to solve the 
performance anomaly by introducing tiny modifications in the 
IEEE~802.11 standard itself, as we do in our solution.

In this context, there exist three main approaches: 
packet fragmentation, contention window adaptation and packet 
aggregation.  
In the following subsections,  we describe briefly each approach 
and we give few relevant examples to illustrate this state of the art. 


\noindent{\bf \em Packet Fragmentation Approach}\\
Packet fragmentation is the first and simplest approach.
Iannone \emph{et Al.} \cite{IF05} propose a solution based on 
a virtual time division scheme that reduces the performance 
anomaly of IEEE~802.11. 
In this solution packets of higher layers are fragmented 
according to the transmission rate at which they are sent at the
802.11 MAC level. 
The packet fragment size is fixed and computed offline.
Simulation results, presented in that work, show that this 
solution reduces performance anomaly while increasing global throughput. 
Nevertheless, the static nature of the proposed solution is efficient
only for stations transmitting at the higher bit rate with a packet
size equal to the MTU on the network. The performance of the network
decreases when only slow hosts are present in the network, due to the
overhead introduced by the high level of fragmentation in small packets. 
A similar approach is proposed by Dunn \emph{et Al.}~\cite{Dun04}, 
but at a higher level. 
The MTU discovery process is used to determine the packet size 
according to the data rate. This solution has the same poor
performance of the previous when only slow hosts are present 
in the network.  

\noindent{\bf \em Contention Window Adaptation Approach}\\
The second category of solution is based on the modification of the
backoff mechanism, in particular changing the contention window ($CW$)
size. 
Heusse \emph{et Al.}~\cite{Heu05} propose a two-step mechanism 
scheme based on the station data rate.  
The first step is a protocol that tries to reach an optimal $CW$ size. 
This optimal value ($CW_{opt}$) is computed according to the number of 
idle slots perceived on the medium by the station. 
Then, in a second step, this $CW_{opt}$ is modified according to the
data rate of the station and the maximum available data rate of the
network. The proposed protocol reduces the performance anomaly 
while improving the throughput. 
The authors show that the main issue of the protocol is the way to 
compute the optimal windows.
The optimal windows values are computed offline according to a fixed 
data rate.
Another problem that can be encountered with this protocol is the 
long convergence time especially when stations are mobile.  


\noindent{\bf \em Packet Aggregation Approach}\\
The third and last category is the packet aggregation approach,
in which our solution is also included.
This type of solution was first introduced by 
Sadeghi \emph{et Al.}~\cite{Sad02}. 
The authors propose an opportunistic media access for multi-rate 
ad hoc networks. The solution is based on the fact that
a station transmitting at high data rate likely to have good channel
condition and thus is allowed to send more than one packet to take
advantage of this favorable channel condition. 
The number of successive packets to transmit is computed according 
to the basic rate of the network. For example if the basic rate is
2Mbps and the channel condition is sensed such that transmission at 
11Mbps is feasible, the sender is granted a channel access time 
sufficient to send $11\%2 = 5 $ packets. 
With this solution, performance anomaly can be solved. However, if
there are only fast stations on the network, short term unfairness
appears.  

The packet aggregation solution is also proposed in the IEEE 802.11e 
standard \cite{802.11e}. In IEEE 802.11e, a transmission
opportunity (TXOP), \emph{i.e.} a maximum channel occupation time, is
granted to every station.  This transmission opportunity is
broadcasted by the base station to every node. 
The computation of TXOP is not really clear in the standard, 
and, as far as we know, it is computed according to the
time needed to send the MTU at the lowest data rate. 
Thus during a TXOP fast stations can aggregate their packets, 
while slow stations can only send one packet. 
The main problem of IEEE 802.11e is that it is centralized. 
Another problem with a static packet aggregation is that the 
performance anomaly is solved on one hand but short time unfairness 
may arise on the other hand. 


To solve the performance anomaly and at the same time this possible
short time unfairness issue, we propose a dynamic packet aggregation
policy. Our solution is different from the other aggregation solutions
because it is not centralized but totally distributed and because it
is not static but totally dynamic. The transmission time is computed
dynamically at each node, according to simple information perceived on
the medium as 
we will describe it on the next section. 
Our approach does not need any additional information except those provided
by IEEE 802.11.

%% file: protocol.tex
The idea of our protocol, called PAS (Performance Anomaly Solution), 
is based on the fact that each station should have the same
transmission time on the radio channel. Therefore, if an emitter
senses a channel occupancy time that is longer than the transmission
time of the packet to be emitted, then it can aggregate packets in
order to get a better channel occupancy time. The aggregation is
realized by spacing the reception of the previous packet's
acknowledgment and the emission of the next packet with a SIFS. 
There are two main mechanisms in PAS: the first one is the medium
sensing that computes the transmission time; the second one is the
packets sending, based on the transmission time computed
previously.

\subsection{Computing the transmission time}
The first mechanism for the computation of the allowed transmission 
time is given in Algorithm \ref{alg:reception}. 
A station always senses the radio medium and maintains the channel
occupancy time. This time is the time during which the channel is
sensed busy due to a transmission, including transmission that can be
only sensed but not decoded ({\em i.e.} in the carrier sensing area). 
The maximum channel occupancy time is maintained by each station in a
variable called $t\_p\_max$. This parameter is set to $0$ after each
successful transmission of the station. This avoids the station to
monopolize the channel after a transmission and improves the
reactivity of the protocol. 
Furthermore, this mechanism allows to reduce the short time 
unfairness that can be introduced when the same node successively
accesses the radio channel.

Note that with this approach, the computed transmission time will
never correspond to the time required for an exchange of packets like
Data-ACK or RTS-CTS-Data-ACK, since this time is deduced from a
continuous signal and will be recomputed as soon as there is a silence
period. Moreover, it is very difficult to determine these
exchanges times since our computation takes into account signals in
the carrier sensing area and that it is not always possible to
distinguish a control packet (RTS, CTS or ACK) from a data packet with
the same transmission time.

\begin{algorithm}[!th]
\caption{ {\textbf Performance Anomaly Solution - Sensing Phase} } \label{alg:reception}
\SetVline
$t\_p\_max$ := 0\;

\Repeat{}{
\If{a signal is sensed at the physical layer}{
$t\_p\_current$ := channel occupancy time of the signal\;
\If{($t\_p\_current > t\_p\_max$)}{$t\_p\_max$ := $t\_p\_current$\;}
\If{(packet type == ACK) and (Dest == me) }
{$t\_p\_max$ := 0;}
}
}
\end{algorithm}

\subsection{Packet emission}

The second mechanism concerns the emission phase and is given in 
Algorithm~\ref{alg:agg}. 
The station can either transmit its packet classically by using the 
medium access mode of IEEE 802.11 or aggregate some of its packets. 
To know whether it can aggregate or not, it uses the parameter 
$t\_p\_max$: if its channel occupancy time is smaller than the 
value of this variable, then it can aggregate. 
In Algorithm~\ref{alg:agg}, $t\_my\_packet$ is the time required to
send the current packet, while $t\_my\_left$ corresponds to
the remaining allowed transmission time. 
The value of this last parameter evolves with time and with the 
packets previously emitted. 
When this value becomes too small, no more aggregation is possible, 
otherwise the medium occupancy time of this station would
become larger than the maximum transmission time sensed on the channel,
which is not fair. 

The boolean variable $sending$ indicates whether the packet to send is
the first packet to be emitted or not. 
If it is the first ($sending$ set to $false$), the packet has to be 
emitted with the classical medium access of IEEE 802.11.
If it belongs to an aggregated packets series ($sending$ set to
$true$), in this case two consecutive packets are only separated 
with a $SIFS$.  

\begin{algorithm}[!b]
\caption{{\textbf Performance Anomaly Solution - Emission Phase}\label{alg:agg}}
\SetVline
$sending$ := false\;
$t\_my\_left := 0$\;
\For{$each\ packet\ to\ send$}{
\If{$t\_my\_left \leq 0$}{ $t\_my\_left := t\_p\_max$\;}
$\alpha = (\lceil\frac{t\_my\_left}{t\_my\_packet}\rceil - \frac{t\_my\_left}{t\_my\_packet}) * t\_my\_packet$\;
$t\_my\_left := t\_my\_left - t\_my\_packet$\;
\If{($sending == true$)}
{\If{($t\_my\_left + \alpha > 0$)}{aggregated\ sending\;}
\Else{$t\_my\_left$ := 0\;
			$sending := false$\;
			classical sending\;
}}
\Else{
{\If{($t\_my\_left + \alpha > 0$)}{$sending := true$\; classical sending\;}
\Else{$t\_my\_left$ := 0\;
			classical sending\;
}}
}}
\end{algorithm}

The parameter $\alpha$ is used to maintain a good overall
throughput. 
Indeed, let consider a scenario with two emitters, one
at 11Mbps and one at 5.5Mbps. These two emitters send packets of the
same size. Due to the physical header overhead (the physical header is
sent at the same rate whatever the emission rate), the time for
transmitting two packets at 11Mbps is a little bit longer than the time for
transmitting one packet at 5.5Mbps. Therefore, without the use of the
variable $\alpha$, the fast station will never aggregate and the
performance anomaly will remain present. By choosing:
\begin{equation}
\alpha = (\lceil\frac{t\_my\_left}{t\_my\_packet}\rceil - \frac{t\_my\_left}{t\_my\_packet}) * t\_my\_packet
\end{equation}
packet aggregation and good aggregated throu\-ghput is ensured, due
to the over-appro\-xima\-tion of the transmission time. 
Note that this parameter is the smallest over-appro\-ximation of the 
transmission time. 
A new value of $\alpha$ is computed at each new packet arrival at the 
MAC layer. 
Thus, we have a real dynamic approach adapted to the current traffic. 
Furthermore, such an approach does not require a specific assumption 
on the packet size.  

If a collision occurs on a packet sent with the classical medium
access of IEEE 802.11, then the collision resolution mechanism of 
IEEE 802.11 is applied. If a collision occurs on a packet sent on an
aggregated packets series, then the transmission is deferred after a
SIFS if $t\_my\_left$ is large enough to send the packet
again. Otherwise if $t\_my\_left$ is too small, the backoff window
size is increased according to the binary exponential backoff scheme
and $sending$ is set to $false$, while $t\_my\_left$ is set to
$0$. 
In the sake of simplicity and due to space constrains,
the collision part is omitted. 

\subsection{Further Improvement}
The transmission time is determined by computing on line the number of packets
that can be emitted and whose total time corresponds to the maximum
channel occupancy perceived on the channel. 
The transmission time of one packet includes the time to transmit the 
packet header. 
Therefore, if a fast station aggregates many small packets,
then a lot of time is lost due to overhead and the overall throughput of
network may not be very good. 
To improve the overall throughput, it is possible to penalize the 
stations that send small packets. 
An easy way to do it is to compute the ratio between packet payload 
and packet header (including acknowledgement), 
we call this ratio $t\_rate$, and to use this parameter to limit
the aggregation. 
In our proposition (PAS), the computation of the next value 
of $t\_my\_left$ is conditioned by the value of $t\_rate$. 
For instance, if $t\_rate<1$,
$t\_my\_left=t\_my\_left- ((1/t\_rate)*t\_my\_packet$). 
At each step this test will reduce the time left for the aggregation 
of a station that sends small packets. 
If at the next step, the packet does not satisfy this test,
$t\_my\_left$ is then computed normally.

In order for to be compatible with all the 802.11 features, it must
work also in presence of RTS/CTS. 
In this case, PAS uses the duration time given in RTS and CTS frames 
to update its maximum occupancy time if this duration time is greater 
than the maximum occupancy time computed previsously. 
The parameter $t\_my\_left$ is still computed like in
Algorithm~\ref{alg:agg}. 
Considering transmission, when $t\_p\_max \geq t\_my\_packet$ 
and $packet_{length}\geq RTS_{thresh}$, then the
exchange is as follow: RTS-CTS-DATA-ACK-SIFS-DATA-\-ACK\ldots. 
The duration time in the RTS and CTS is the duration for only one packet
transmission. There are two reasons to not put the value of
$t\_p\_max$ in the duration field of the RTS and CTS frames: {\it i)}
since the number of packets in the LL queue is not known {\em a priori}
when a RTS is sent, it is possible that the emitter will not use its
whole transmission time, which will unnecessarily stop some potential
emitters; {\it ii)} reactivity is improved. If we assume
two fast stations and one slow station, the two fast stations may
aggregate their packets based on the transmission time of the slow
station. If the slow station stops emitting, the two fast stations
will maintain their aggregation because the duration field remains the
same for these two stations.

With PAS, collisions, when RTS/CTS mechanism is used, are solved in 
the following way. If a collision occurs on a RTS, the RTS is retransmitted
according to IEEE 802.11, {\textit i.e.} after a backoff window
incrementation. When a collision occurs on the data, the data packet
is sent after a SIFS, if $t\_my\_left$ is large enough to send the
packet again. If $t\_my\_left$ is not large enough, then a RTS is sent
after a backoff window incrementation.

%% file: analytical.tex
In this section, we investigate the efficiency and the
fairness achieved by PAS. Tan \emph{et Al.}~\cite{TG04} have proposed
the notion of time-based fairness that gives to each node an approximately
equal occupancy of the channel. They show that a
mechanism that provides a time-based fairness is more
efficient than a mechanism that is fair in the medium access. 
The solution they propose\footnote{The work has not been described in
Section~\ref{sec:art}, since the solution is also considered at upper layers
and not only at the MAC layer.} takes into account the time
required for the exchange data-ACK in the computation of the
transmission time, whereas PAS is based on the maximum channel
occupancy that can never be such an exchange. In this section, we show
that PAS is more efficient than solutions based on data-ACK
exchanges and we study the time-based fairness of PAS.

\subsection{Efficiency}
The time transmission in our protocol is based on packet time and not
on the time required for an exchange. 
An exchange time can be defined as  
$T\_ex = t\_my\_packet + T\_SIFS + T\_PHY + T\_ACK $, where $T\_SFIS$ is the
duration of a SIFS, $T\_PHY$ is the duration of the PHY header and
$T\_ACK$ is the time duration of an ACK. 
By $t\_p\_max$ we denote the maximum channel occupancy time, 
by $t\_my\_packet$ the time required to transmit the packet, 
and by $T\_ack$ the sum of $ T\_SIFS + T\_PHY + T\_ACK$. 
We assume that $T\_ack$ is independent from the data rate at which a node
transmits and is a constant. 
We also assume as scenario two stations within communication range
from each other (one fast station and one
slow station) that use the same packet length.
 The maximum aggregate throughput is obtained when the
fast station aggregate as much packets as possible, on the basis of
the transmission time of the slow station.
The number of packets sent by the fast station with PAS is given by: 
\begin{equation}
n_{a}=\frac{t\_p\_max}{t\_my\_packet}
\label{eq:1}
\end{equation}
while the number of packets sent by the fast station using the
exchange time for the aggregation, like in the work of Tan \emph{et
Al.} \cite{TG04}, is given by:
\begin{equation}
n_{et}=\frac{t\_p\_max + T\_ack}{t\_my\_packet + T\_ack}
\label{eq:2}
\end{equation}
We have $t\_my\_packet \leq t\_p\_max$. Thus, with these assumptions:
\begin{equation}
n_{a} \geq n_{et}
\label{eq:3}
\end{equation}
Therefore, each time the slow station sends a packet, the fast
station, in its next transmission, will aggregate more packets with
PAS than with the solution proposed by Tan \emph{et Al.} \cite{TG04},
showing the higher efficiency of PAS.

\subsection{Fairness}

In this section, we investigate the time-based fairness as discussed
by Tan \emph{et Al.}. 
In the sake of simplicity,  in this analysis  we
assume that each node uses the same packet length $L$ (in bytes). 
We also assume that 
$T_{i}$ with $i=1,2,5.5,11$ is the time needed to transmit a packet at
data rate $i$Mbps. $T_{i}$ includes the transport layer header, the
network layer header, the MAC layer header and PHY layer header. We
can easily compute the time used by a station transmitting at a data
rate $i$ as:
\begin{equation}
Agg_{i} = n_{a_{i}}\times(T_{i} + T\_ack) + (n_{a_{i}}-1)\times T\_SIFS
\label{eq:aggregation}
\end{equation}
$Agg_i$ is the time required for the aggregated transmission of a node
transmitting at data rate $i$, where $n_{a_{i}} =
t\_p\_max/T_{i}$. From the medium point of view, the time proportion
used for an aggregated transmission of one node is: 
\begin{equation}
Occ_{i} = \frac{Agg_{i}}{\sum_j (Agg_{j}\times N_j) + N*DIFS}
\label{eq:occupation}
\end{equation}
where $N_j$ is the number of stations transmitting at a data rate $j$,
with $\sum_j N_j = N$. We assume here that the probability to access
the medium is the same for all the nodes and that during a time
interval, each node has accessed the medium exactly once. The number of
packets sent by a node transmitting at a data rate $i$, in a time
interval $t$, is: 
\begin{equation}
NBp_{i} = \frac{n_{a_{i}}}{\sum_{j} (Agg_{j}\times N_j) + N\times(DIFS+Avg_{bckf})}\times t
\label{eq:packet}
\end{equation}
where $Avg_{bckf}$ is the average backoff. 
We can thus derive the average throughput in bps of a station
transmitting at a data rate $i$ with the following equation:
\begin{equation}
TH_{i} =  NBp_{i}\times L\times 8
\label{eq:throughput}
\end{equation}
All the above results can be applied with different packet sizes, the
main parameter to know is $t\_p\_max$. In this analysis, we assume
that stations access to the medium in a TDMA mode, {\em i.e.} one
station after the other. This assumption is legitimate due the
fair access provided by the backoff scheme implemented in the DCF
of IEEE 802.11. However, we will see, in the following section, that
there are some small differences between the analytical results and
the simulation results and that these differences come from this
assumption. Indeed, IEEE 802.11 does not provide a perfect TDMA
scheduling in the short-term.

\begin{figure}[tpb]
\centering
\includegraphics[scale=0.55]{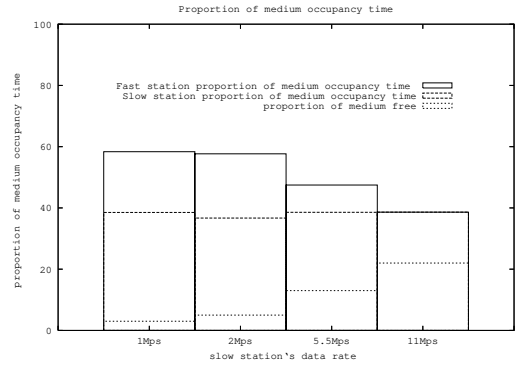}
\caption{Proportion of medium occupancy time for two stations}
\label{fig:time_proportion}
\end{figure}

Figure~\ref{fig:time_proportion} shows, for two stations, the
proportion of medium occupancy time. One of the two stations transmits
at 11Mbps while the other transmits at 1, 2, 5.5, or 11Mbps (on the
x-axis, $i$Mbps indicates that one station emits at $i$Mbps while the
other emits at 11Mbps). Packet size is equal to 1000 bytes. For each
$i$, this figure gives the proportions of medium occupancy time of the
fast station (11Mbps) and of the slow station ($i$Mbps) and the time
proportion when the medium is free.  We can see that the fast station
gets a larger proportion of medium occupancy than the slow station and
that the proportion of each station is not $50\%$ as it should be with a
perfect time-based fairness. This difference may be easily explained
by the fact that the allowed transmission time computed with PAS does
not take into account the acknowledgments that consume transmission
time. We can also see from this figure that the higher the data rate
of the slow station, the higher the proportion of medium free. This is
due to the proportion between the backoff time and the medium
occupancy time that increases. 

\begin{table}[!tp]
\begin{center}
\begin{tabular}{|c|c|c|c|}
\hline
                                          & Th. (kbps) & Pkt nb. (/s)&Index\\
\hline
    5.5Mbps &   1547.2   &  193.4   & 0.98\\
    11Mbps  &   3095.2   &  386.9   &\\

\hline
    2Mbps 	&   624.8    &  78.1    & 0.93 \\
    11Mbps  &   3749.6   &  468.7   &\\
\hline
    1Mbps 	&   344.8    &  43.1    &	0.92\\
    11Mbps  &   3791.2   &  473.9   &\\
\hline
\end{tabular}
\caption{PAS: analytical results\label{tab:analytical}}
\end{center} 
\end{table}

Table~\ref{tab:analytical} shows the throughput obtained by
Equation~\ref{eq:throughput}. 
We included the Jain fairness index \cite{Jai99} to evaluate the 
fairness of our solution. The Jain index
is defined as $\frac{(\sum_i{r_i/r^*_i})^2}{n \sum_i{(r_i/r^*_i)^2}}$,
where $r_i$ is the rate achieved on flow $i$, $n$ is
the number of flows, and $r^*_i$ is the reference rate on flow $i$. 
As reference rate we use the one defined by Tan \emph{et Al.}. 
This rate $r^*_i$ is computed as if all the flows in the wireless 
networks were emitted at the same data rate as flow $i$. For
example, if we consider two nodes transmitting at $11$ (flow $1$) and
$1$Mbps (flow $2$). Then $r^*_1$ will be the throughput of flow $1$
if flow $2$ is transmitted at $11$Mbps. In the same way, $r^*_2$ will be
the throughput of flow $2$ if flow $1$ is transmitted at $1$Mbps. 
The value of $r^*_i$ is the throughput value when the medium occupancy
time is equal for all nodes. This is the reason why the index computed in 
table~\ref{tab:analytical} are not equal to $1$.

%% file: simulation.tex
\label{sec:simulations}

The NS-2 simulator \cite{NS2} is used to evaluate PAS, which is
coded as an independent MAC. Multi-rate features are also  
added to the simulator, in order to reflect the IEEE 802.11
modulations. 
All the studies listed below are done in steady state condition. 
In order to reduce the simulation time and to better evaluate the 
protocol, ARP and routing protocol exchanges are disabled.
In all simulations UDP saturated traffic is used. 
If not differently specified, each packet contains 1000 bytes of data. 
Nevertheless, we also developed a module to generate packets of a random size, 
uniformly distributed in a specific interval. 

\subsection{Model validation}
In order to validate the improvements to NS-2 and the code of our 
proposal, we first simulate two pairs of station transmitting at
11Mbps with 1000 bytes of data. 
In this simulation, no aggregation is done because the maximum 
occupancy time perceived by each node is equal to the time required 
to send a packet. 
In this specific case, the throughput of 802.11 and PAS should be the 
same. 
This is confirmed by the results presented in
Table~\ref{tab:validation}, which includes the theoretical throughput
derived in Section~\ref{sec:theo}, in order to show the accuracy of our model.

\begin{table}[!tbp]
\begin{center}
\begin{tabular}{|l|c|c|c|}
\hline
																	& 				& Th. (kbps) & Conf. Int. (0.05)                            \\
\hline                                                                                                                                                                           
\multirow{3}{*}{802.11} 		  		& 11Mbps	&	2747.04 & [2731.35 ; 2762.72]	 \\
                                	&	11Mbps	&	2752.80 & [2736.80 ; 2768.81]	 \\
                                	&	Total		&	5499.84 & [5491.02 ; 5508.66]	 \\
\cline{3-4}		&	Index		&\multicolumn{2}{c|}{0.99999}\\
\hline                                                                       
\multirow{3}{*}{PAS} 							& 11Mbps	&	2740.61 & [2726.91 ; 2754.30]	 \\
																	&	11Mbps	&	2753.71 & [2740.51 ; 2766.92]	 \\
																	&	Total		&	5494.32 & [5485.78 ; 5502.86]	 \\
\cline{3-4}		&	Index		&\multicolumn{2}{c|}{0.99999}\\
\hline
\multirow{3}{*}{Theoretical} 
							&	11Mbps		&\multicolumn{2}{c|}{2802.5919 (kbps)}\\
							&	11Mbps		&\multicolumn{2}{c|}{2802.5919 (kbps)}\\
							&	Total			&\multicolumn{2}{c|}{5605.1839 (kbps)}\\
\hline
\end{tabular}
\caption{Model validation\label{tab:validation}}
\end{center}
\end{table}  

\subsection{Basic simulations}
This section contains the first simulation results of PAS. 
The simulation carried out is based on the classical scenario 
where two stations transmit packets of 1000 bytes, one at xMbps 
(x equal to 1, 2 or 5.5) and the other at 11Mbps.
Tables~\ref{tab:normal_debit_1},~\ref{tab:normal_debit_2}
and~\ref{tab:normal_debit_3} give the simulation results in this
scenario. 
In these tables, we give the achieved throughput of each
station, the achieved overall throughput, the number of sent packets
by each station and in total, as well as the Jain fairness index, 
introduced in Section~\ref{sec:theo}. 

One can see from these tables that the aggregate throughput of PAS is
always greater than 802.11, thus PAS is more efficient. 
It can also be observed that when using PAS, the number of packets 
and the throughput of the fast station remain almost the same, 
independently of the rate used by the slow station. 
This is because the time occupation is roughly divided by 2 between  
the fast station and the slow station. The fairness index shows that PAS
achieves a very good fairness in terms of medium occupancy in these
scenarios. 

\begin{table*}[!htbp]
\begin{center}
\begin{tabular}{|l|c|c|c|c|c|c|}
\hline
																	& 				& Th. (kbps) & Conf. Int. (0.05)                            & Packets/s & Conf. Int. (0.05)& Fairness index\\
\hline                                                                                                                                                                           
\multirow{3}{*}{802.11} 		  		& 5.5Mbps	&		2157.02							& [2147.86 ; 2166.19]          &		258.79								& [257.34 ; 260.24]&\multirow{3}{*}{0.9556825}\\
                                	&	11Mbps	&		2111.78 						&	[2099.96 ; 2123.61]          &	  264.34								& [263.21 ; 265.46]&\\
                                	&	Total		&		4268.81							&	[4260.53 ; 4277.10]          &		523.13								&	[522.12 ; 524.15]&\\
\hline                                                                                                                                               
\multirow{3}{*}{PAS} 							& 5.5Mbps	&		1769.89							& [1761.23 ; 1778.54]          &		216.89								& [215.83 ; 217.95]&\multirow{3}{*}{0.9978824}\\
																	&	11Mbps	&		2943.07 						&	[2927.82 ; 2958.32]          &		360.67								&	[358.80 ; 362.53]&\\
																	&	Total		&		4712.96							&	[4703.02 ; 4722.91]          &		577.56								&	[576.35 ; 578.78]&\\
\hline
\end{tabular}
\caption{Performance anomaly results (throughput and  number of packets)\label{tab:normal_debit_1}}
\end{center}
\end{table*}

\begin{table*}[!htbp]
\begin{center}
\begin{tabular}{|l|c|c|c|c|c|c|}
\hline
																	& 				& Th. (kbps) & Conf. Int. (0.05)                            & Packets/s & Conf. Int. (0.05)& Fairness index\\
\hline                                                                                                                                                                           
\multirow{3}{*}{802.11} 					& 2Mbps		&		1240.93							& [1236.03 ; 1245.84]          &		152.07								& [151.47 ; 152.67]&\multirow{3}{*}{0.7676374}\\
                                	&	11Mbps	&		1219.97 						&	[1203.54 ; 1236.39]          &	  149.50								& [147.49 ; 151.51]&\\
                                	&	Total		&		2460.91							&	[2447.07 ; 2474.74]          &		301.58								&	[299.88 ; 303.27]&\\
\hline                                                                                                                                               
\multirow{3}{*}{PAS} 							& 2Mbps		&		816.51							& [811.19 ; 821.83]       	   &		100.06								& [99.41 ; 100.71]&\multirow{3}{*}{0.9976767}\\
																	&	11Mbps	&		3046.88 						&	[3023.13 ; 3070.62]          &		373.39								&	[370.48 ; 376.30]&\\
																	&	Total		&		3863.39							&	[3843.14 ; 3883.64]          &		473.45								&	[470.97 ; 475.93]&\\
\hline
\end{tabular}
\caption{Performance anomaly results (throughput and  number of packets)\label{tab:normal_debit_2}}
\end{center}
\end{table*}  

\begin{table*}[!htbp]
\begin{center}
\begin{tabular}{|l|c|c|c|c|c|c|}
\hline
																	& 				& Th. (kbps) & Conf. Int. (0.05)                            & Packets/s & Conf. Int. (0.05)& Fairness index\\
\hline                                                                                                                                                                           
\multirow{3}{*}{802.11} 					& 1Mbps		&		  740.60 & [737.31 ;   743.88]     &	90.76 & [90.36 ;    91.16]		&\multirow{3}{*}{0.6497743}\\
                                	&	11Mbps	&		  726.45 & [710.65 ;   742.24]     &	89.03 & [87.09 ;    90.96]  	&\\
                                	&	Total		&		  1467.04 & [1452.14 ;  1481.95]    &	179.78 & [177.96 ;   181.61]	&\\
\hline                                                           
\multirow{3}{*}{PAS} 							& 1Mbps		&		  461.81 & [457.45 ;   466.18]	   &	56.59 & [56.06 ;    57.13]		&\multirow{3}{*}{0.9999946}\\
																	&	11Mbps	&		  2941.32 & [2910.81 ;  2971.83]    &	360.46 & [356.72 ;   364.19]	&\\
																	&	Total		&		  3403.13 & [3375.51 ;  3430.75]    &	417.05 & [413.67 ;   420.44]	&\\
\hline
\end{tabular}
\caption{Performance anomaly results (throughput and  number of packets)\label{tab:normal_debit_3}}
\end{center}
\end{table*}  

The difference between the theoretical results 
(Table~\ref{tab:analytical}) and the simulation 
results can be explained by the backoff algorithm present in the 
IEEE 802.11 MAC. Indeed, the backoff algorithm does
not provide a TDMA-like access to the medium. When there are only two
stations, each station can access successively the medium. 
In the case of PAS, the fast station will first aggregate its 
packets during its transmission time and when its transmission time 
elapses, it will send its packets classically with IEEE 802.11 if it 
accesses successively to the medium. 
Therefore, this feature of PAS reduces the throughput of the fast 
station because it does not always aggregate its packets. 
This reduction can be worsened when the slow station sends
also successive packets. The difference between the analytical results
and the simulation results increases when the difference in the data
rate of the two stations increases.

\begin{table*}[!htbp]
\begin{center}
\begin{tabular}{|l|c|c|c|c|c|c|}
\hline
																	& 				& Th. (kbps) & Conf. Int. (0.05)                            & Packets/s & Conf. Int. (0.05)& Fairness index\\
\hline                                                                                                                                                                           
\multirow{5}{*}{802.11} 					& 1Mbps		&		423.08 & [415.67 ;   430.49]      &	51.85 & [50.94 ;    52.76]&\multirow{5}{*}{0.6598870} \\
                                	&	2Mbps		&		413.68 & [403.86 ;   423.50]      &	50.70 & [49.49 ;    51.90]&  \\
                                	&	5.5Mbps	&		401.80 & [389.96 ;   413.65]      &	49.24 & [47.79 ;    50.69]&  \\
                                	&	11Mbps	&		392.09 & [379.93 ;   404.26]      &	48.05 & [46.56 ;    49.54]&  \\
                                	&	Total		&		1630.66 & [1614.28 ;  1647.04]     & 199.84 & [197.83 ;   201.84]&  \\
\hline                                                           											 
\multirow{5}{*}{PAS} 							& 1Mbps		&		 236.02 & [230.10 ;   241.94] 	  &	28.92  & [28.20 ;    29.65]&\multirow{3}{*}{0.9972993}\\
																	&	2Mbps		&		 376.81 & [366.19 ;   387.42]     &	46.18  & [44.88 ;    47.48]&	\\
																	&	5.5Mbps	&		 943.25 & [917.63 ;   968.88]     &	115.59 & [112.45 ;   118.74]&	\\
																	&	11Mbps	&		 1499.68 & [1453.82 ;  1545.55]    &	183.78 & [178.16 ;   189.41]&	\\
																	&	Total		&		 3055.77 & [3021.34 ;  3090.19]    &	374.48 & [370.26 ;   378.70]&	\\
\hline
\end{tabular}
\caption{Performance anomaly results (throughput and  number of packets)\label{tab:normal_debit_4}}
\end{center}
\end{table*}  

\begin{table*}[!htbp]
\begin{center}
\begin{tabular}{|l|c|c|c|c|c|c|}
\hline
																	& 				& Th. (kbps) & Conf. Int. (0.05)                            & Packets/s & Conf. Int. (0.05)& Fairness index\\
\hline                                                                                                                                                                           
\multirow{5}{*}{802.11} 					& 1Mbps		&		260.71 & [251.58 ;   269.83]    &	31.95 & [30.83 ;    33.07]&\multirow{5}{*}{0.8222611}\\
                                	&	1Mbps		&		253.68 & [244.85 ;   262.52]    &	31.09 & [30.01 ;    32.17]&\\
                                	&	1Mbps		&		259.36 & [250.78 ;   267.95]    &	31.78 & [30.73 ;    32.84]&\\
                                	&	11Mbps	&		267.21 & [256.25 ;   278.18]    &	32.75 & [31.40 ;    34.09]&\\
                                	&	Total		&		1040.97 & [1030.81 ;  1051.13]   & 127.57 & [126.32 ;   128.81]&\\
\hline                                          			
\multirow{5}{*}{PAS} 							& 1Mbps		&		213.50 & [206.55 ;   220.46]	  &	26.16 & [25.31 ;    27.02]&\multirow{5}{*}{0.9980227}\\
																	&	1Mbps		&		210.30 & [202.72 ;   217.88]    &	25.77 & [24.84 ;    26.70]&\\
																	&	1Mbps		&		202.45 & [193.29 ;   211.61]    &	24.81 & [23.69 ;    25.93]&\\
																	&	11Mbps	&		1540.59 & [1488.93 ;  1592.24]   & 188.80 & [182.47 ;   195.13]&\\
																	&	Total		&		2166.84 & [2120.97 ;  2212.71]   & 265.54 & [259.92 ;   271.17]&\\
\hline
\end{tabular}
\caption{Performance anomaly results (throughput and  number of packets)\label{tab:normal_debit_5}}
\end{center}
\end{table*}

\begin{table*}[!htbp]
\begin{center}
\begin{tabular}{|l|c|c|c|c|c|c|c|}
\hline
																	& 				& Th. (kbps) & Conf. Int. (0.05)                            & Packets/s & Conf. Int. (0.05)& Fairness index\\
\hline                                                                                                                                                                           
\multirow{5}{*}{802.11}			 			& 1Mbps		&	 330.53 & [320.45 ;   340.61]	   &40.51 & [39.27 ;    41.74]&\multirow{5}{*}{0.6822219}	\\
                                	&	1Mbps		&	 345.51 & [336.32 ;   354.70]	   &42.34 & [41.22 ;    43.47]&	\\
                                	&	5.5Mbps	&	 341.89 & [328.66 ;   355.13]	   &41.90 & [40.28 ;    43.52]&	\\
                                	&	11Mbps	&	 332.60 & [319.99 ;   345.20]	   &40.76 & [39.21 ;    42.30]&	\\
                                	&	Total		&	 1350.53 & [1335.64 ;  1365.43]	 &165.51 & [163.68 ;   167.33]&	\\
\hline                                          
\multirow{5}{*}{PAS} 							& 1Mbps		&		208.13 & [201.54 ;   214.72]   &	25.51 & [24.70 ;    26.31]&\multirow{5}{*}{0.9991965}\\
																	&	1Mbps		&		214.23 & [208.10 ;   220.35]   &	26.25 & [25.50 ;    27.00]&\\
																	&	5.5Mbps	&		949.87 & [922.42 ;   977.31]   &	116.41 & [113.04 ;   119.77]&\\
																	&	11Mbps	&	 1510.32 & [1465.07 ;  1555.58]	 &	185.09 & [179.54 ;   190.63]&\\
																	&	Total		&		2882.55 & [2848.88 ;  2916.21]  &	353.25 & [349.13 ;   357.38]&\\
\hline
\end{tabular}
\caption{Performance anomaly results (throughput and  number of packets)\label{tab:normal_debit_6}}
\end{center}
\end{table*}  

Tables~\ref{tab:normal_debit_4},~\ref{tab:normal_debit_5}
and~\ref{tab:normal_debit_6} show the simulation results with four
stations transmitting respectively at $\{1,2,5.5,11\}$Mbps, at
$\{1,1,1,11\}$Mbps and at $\{1,1,5.5,11\}$Mbps. From these results, one
can see that the aggregate throughput of PAS is always greater than
the aggregate throughput of 802.11. 
The throughput and the number of packets for the fast stations 
(especially at 11Mbps) with PAS remain almost the same in the 
different tables. 
This is because the time accorded to  each station to send its packets 
is based on the slowest packet time transmission. 
The fairness index also shows that PAS is fair in terms of medium 
occupancy.  



\subsection{Reactivity}
\begin{figure}[tpb]
\centering
\includegraphics[scale=0.5]{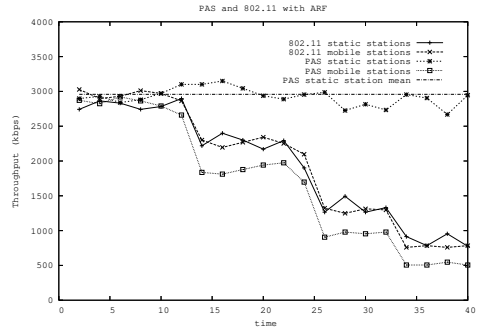}
\caption{PAS implemented with ARF}
\label{fig:ARF}
\end{figure}

A way to test the reactivity of PAS is to introduce the
well-known Auto-Rate Fallback (ARF) mechanism used by wireless stations to
adapt their transmission rate to the channel conditions. 
We have implemented the ARF mechanism to see the behavior of PAS when
the transmission rates of stations vary in time. 
The simulation is done using two emitters with one station moving away from the
other. Figure~\ref{fig:ARF} shows the simulation results with PAS and
802.11. We can see from this figure that when using PAS, the
throughput of the fast station remains constant, while the throughput of
the moving station decreases. With IEEE 802.11, the throughput of the
two emitters decreases.

\subsection{Delay}

In this section we present a simulation of 20 seconds with 2 emitters:
one with a data rate of 11Mbps and the other with a data rate of
1Mbps. During this simulation we compute the inter-burst time. An
inter-burst time is defined as the time between the end of
a burst and the beginning of another burst from the same station. 
For the station transmitting at the lower data rate a burst consists 
always in a single packet. 
For the station transmitting at the higher data rate, a burst
can be either a real packet burst (several aggregated packets) or a
single packet if the wireless station accesses the medium immediately
after a burst. 

Table~\ref{tab:burst} gives the number of sent bursts and the average
inter-burst time for the two stations. One can see that IEEE 802.11 
provides a fair access to the medium, since the number of bursts 
for the slow and the fast stations is nearly the same. 
The table also shows that the average inter-burst time
is close to the packet transmission time of the slow 
station~($8576\mu s$).

\begin{table}[tbp]
\centering
\begin{tabular}{|l|c|c|}
\hline
							& 	Nb bursts	& Avg inter-burst \\
\hline
FAST 					& 		5911		&				9867.70$\mu$s		\\
\hline
SLOW					&			6004		&				8776.46$\mu$s		\\
\hline
\end{tabular}
\caption{PAS: delay\label{tab:burst}}
\end{table}

\begin{figure}[!b]
\centering
\includegraphics[scale=0.5]{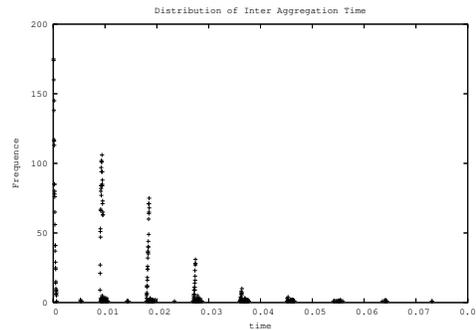}
\caption{Inter-burst time distribution for the fast station}
\label{fig:11freq}
\end{figure}

Figure~\ref{fig:11freq} shows the inter-burst time distribution for
the fast station. One can easily see that the medium access provided by
the backoff algorithm is not really a TDMA-like access  due to the
peak close to 0 in the figure. We can also see from this figure that
the presence of successive peaks shows that 
the slow station can send many successive packets. This confirms what
we claim in Section~\ref{sec:theo} about the difference between
simulation and analytical results. In this figure the difference (in
time) between two peaks is close to the packet duration of the slow
station. 

\begin{figure}[!b]
\centering
\includegraphics[scale=0.5]{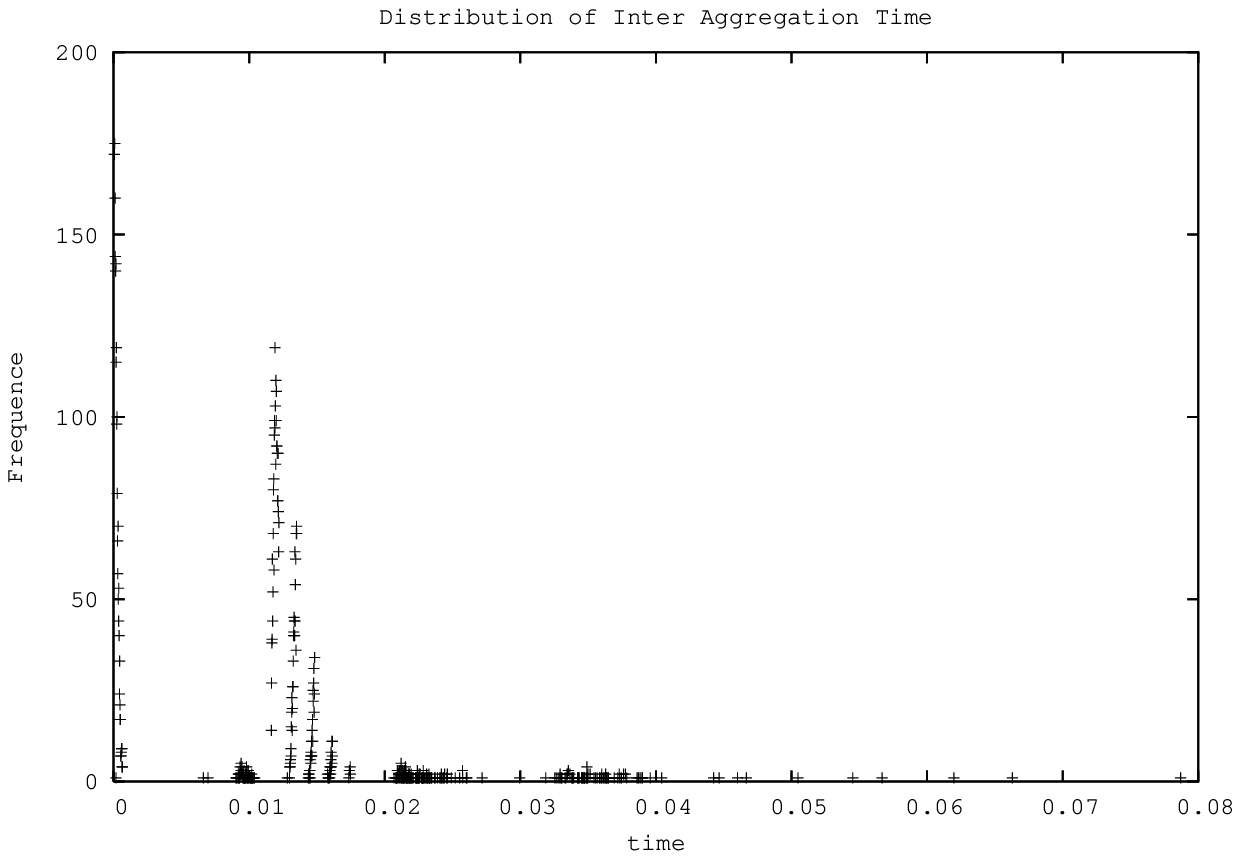}
\caption{Inter-burt time distribution for the slow station}
\label{fig:1freq}
\end{figure}

Figure~\ref{fig:1freq} shows the inter-burst time distribution for
the slow station. One can see from the figure that the average inter-burst
time is close to the time needed by the fast station to transmit
aggregated packets. The distribution presented in his figure is 
completely different from the one presented in previous figure 
(Figure~\ref{fig:11freq}). 
The reason is that even if the fast station can send successive packets, it
is just for the transmission of a single packet and not for a
burst. This also explains that the average inter-burst time of the slow
station is smaller than the one of the fast station.  

In both figures (Figure~\ref{fig:11freq} and
Figure~\ref{fig:1freq}), the points close to $0$ means that there is a
considerable number of packets that are send successively with the
backoff algorithm of IEEE 802.11. Such a feature reduces the
performances of PAS. 

\subsection{Effect of $\alpha$}
In this section, we investigate the effect of the $\alpha$ parameter 
on the performance of PAS. 
We simulate two emitters transmitting 1000 bytes of data at 11Mbps 
and at 5.5Mbps. 
The simulation is carried out with and without the use of $\alpha$. 
One can see from Table~\ref{tab:alpha} that in this specific
simulation, when $\alpha$ is not used, there is no aggregation. 
Indeed, in this case the condition $t\_my\_left - t\_my\_packet > 0$ 
never holds for the fast station, thus it does not
perform any aggregation.
\begin{table}[tbp] 
\centering
\begin{tabular}{|l|c|c|c|}
\hline
																	& 				& Th. (kbps) & Conf. Int. (0.05)\\
\hline
\multirow{3}{*}{PAS w/o $\alpha$} 		& 5.5Mbps	&		2147.31							& [2137.62 ; 2157.01]\\
                                		&	11Mbps	&		2131.51 							&	[2119.42 ; 2143.60]\\
                                		&	Total		&		4278.83							&	[4269.92 ; 4287.74]\\
\cline{3-4}		&	Index		&\multicolumn{2}{c|}{0.9582439}\\
\hline
\multirow{3}{*}{PAS} 					& 5.5Mbps	&		1769.89							& [1761.23 ; 1778.54]\\
															&	11Mbps	&		2943.07 						&	[2927.82 ; 2958.32]\\
															&	Total		&		4712.96							&	[4703.02 ; 4722.91]\\
\cline{3-4}		&	Index		&\multicolumn{2}{c|}{0.9978824}\\
\hline
\end{tabular}
\caption{The influence of $\alpha$ on performances\label{tab:alpha}}
\end{table} 

We have also simulated a scenario with four emitters, respectively at
1, 2, 5.5 and 11~Mbps. 
From Table~\ref{tab:alpha} and Table~\ref{tab:alpha_} we can see
that $\alpha$ increases fairness and efficiency. 
Indeed, when $\alpha$ is used, the proportion of medium occupancy for 
the fast stations is increased. 

\begin{table}[!bp]
\centering
\begin{tabular}{|l|c|c|c|}
\hline
																	& 				& Th. (kbps) & Conf. Int. (0.05) \\
\hline                                                                                                                                                                           
 							& 1Mbps		&		278.47 & [271.66 ;   285.28]   	\\
             	&	2Mbps		&		283.95 & [274.67 ;   293.23]   	  \\
PAS          	&	5.5Mbps	&		880.56 & [857.80 ;   903.32]   	  \\
w/o $\alpha$	&	11Mbps	&		1484.19 & [1438.28 ;  1530.10]  	  \\
             	&	Total		&		2927.17 & [2893.26 ;  2961.08]  	  \\
\cline{3-4}		&	Index		&\multicolumn{2}{c|}{0.9804155}\\
\hline                                                       
 							& 1Mbps		&		 236.02 & [230.10 ;   241.94] 	 \\
PAS	 					&	2Mbps		&		 376.81 & [366.19 ;   387.42]    	\\
							&	5.5Mbps	&		 943.25 & [917.63 ;   968.88]    	\\
							&	11Mbps	&		 1499.68 & [1453.82 ;  1545.55]  	\\
							&	Total		&		 3055.77 & [3021.34 ;  3090.19]  	\\
\cline{3-4}		&	Index		&\multicolumn{2}{c|}{0.9972993}\\
\hline
\end{tabular}
\caption{The influence of $\alpha$ on performances\label{tab:alpha_}}
\end{table}  

\subsection{Effect of $t\_rate$}

Another important parameter of PAS is $t\_rate$. This parameter
controls the time left for an aggregated transmission. 
It increases or reduces the aggregated transmission time, depending 
on the ratio between payload and the header. 
Table~\ref{tab:ratio} gives the results of simulation runs with two 
emitters, one transmitting at 11Mbps with packets of 100 bytes length, the
other transmitting at 5.5Mbps with packets of 1000 bytes length. 
One can see from this table that $t\_rate$ improves the global 
throughput of the network, but this overall throughput is smaller than 
in the case of IEEE 802.11. 
There are several possibilities to improve the use of $t\_rate$. 
For instance, if $t\_rate \leq 1$, setting $t\_my\_left$ to $0$ will 
stop the aggregated sending if a small packet was sent. 
The problem by using this scheme is that when a small packet from 
upper layer arrives (such as ACK from TCP protocol), it always 
penalizes the wireless station when it gains the access to the medium.  

\begin{table}[tbp]
\centering
\begin{tabular}{|l|c|c|c|}
\hline
																								& 				& Th.(kbps) & Conf. Int\\
\hline
\multirow{3}{*}{802.11} 												& 11Mbps	&	308.37 & [299.76 ;  316.98]\\
																								&	5.5Mbps	&	2631.15 & [2586.04 ; 2676.25]\\
																								&	Total		&	2939.52 & [2898.72 ; 2980.32]\\					
\cline{3-4}		&	Index		&\multicolumn{2}{c|}{0.8140598}\\
\hline
\multirow{3}{*}{PAS} 														& 11Mbps	&	458.99 & [446.99 ;  470.98]\\
																								&	5.5Mbps	&	2344.64 & [2313.36 ; 2375.92]\\
																								&	Total		&	2803.63 & [2778.27 ; 2828.98]\\
\cline{3-4}		&	Index		&\multicolumn{2}{c|}{0.9363749}\\
\hline
\multirow{2}{*}{PAS w/o } 											& 11Mbps	&	816.43 & [801.25 ;  831.60]\\
																								&	5.5Mbps	&	1668.27 & [1629.82 ; 1706.71]\\
		$t\_rate$																		&	Total		&	2484.69 & [2456.72 ; 2512.66]\\
\cline{3-4}		&	Index		&\multicolumn{2}{c|}{0.9636280}\\
\hline
\end{tabular}
\caption{The influence of $t\_rate$ on performances\label{tab:ratio}}
\end{table}  

\begin{figure}[!b]
\centering
\includegraphics[scale=0.5]{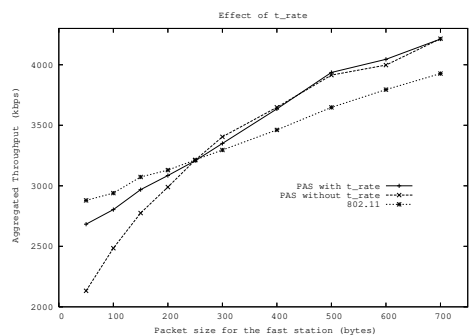}
\caption{Aggregated throughput depending on the packet size\label{fig:agg}}
\end{figure}

\begin{figure}[!b]
\centering
\includegraphics[scale=0.5]{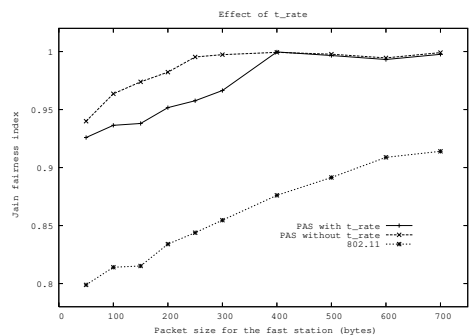}
\caption{Fairness index depending on the packet size\label{fig:index}}
\end{figure}

One can see from Table~\ref{tab:ratio} that $t\_rate$ has a negative
impact on fairness. This because the $t\_rate$ is used to reduce the
aggregation time. In this particular scenario, it appears that there
is a tradeoff between fairness and efficiency. We argue that PAS 
provides this good tradeoff, as Figure~\ref{fig:agg} and 
Figure~\ref{fig:index} confirm. 
One can see from these figures that when using the $t\_rate$, 
PAS is not as efficient as IEEE 802.11 for small values of $t\_rate$, 
however, the aggregated throughput of the two solutions are 
close (Fig.~\ref{fig:agg}). 
Furthermore, for small values of $t\_rate$, the fairness index 
of PAS using $t\_rate$ is lower than the fairness index of PAS not 
using $t\_rate$, however, they are very close
(Fig.~\ref{fig:index}). 

\subsection{Using dynamic packet sizes}
In this section we have tested our protocol with different packet
sizes. 
Packets are generated at each node with a uniform distribution
between 550 bytes and 1450 bytes. Table~\ref{tab:diff_1} shows the
variation of $t\_p\_max$ and $t\_my\_packet$ during the simulation. 
One can see from this table that the difference between the maximum values
and the minimum values of $t\_p\_max$ and $t\_my\_packet$ may be high. 


\begin{table}[!tbp] 
\centering
\begin{tabular}{|l|c|c|c|}
\hline
																	& 				& $t\_p\_max\ (\mu s)$ 					& $t\_my\_packet\ (\mu s)$		 \\
\hline
\multirow{2}{*}{PAS} 					& 5.5Mbps	&		248 - 954							& 320 - 1716\\
															&	11Mbps	&		248 - 1716 						&	285 - 954\\
\hline
\end{tabular}
\caption{PAS with different packet sizes\label{tab:diff_1}}
\end{table} 


Table~\ref{tab:diff} shows the average throughput obtained in previous
simulations. 
One can see that PAS is efficient and fair when using a uniform
distribution for the packet size. This behavior of PAS is possible
because the number of packets to aggregate is not known {\em a priori}
and is computed dynamically at the arrival of each new packet.    

\begin{table}[tbp] 
\centering
\begin{tabular}{|l|c|c|c|}
\hline
																	& 				& Th. (kbps) & Conf. Int. (0.05)\\
\hline
\multirow{3}{1.2cm}{802.11} 			& 5.5Mbps	&		2075.67							& [2065.93 ; 2085.41]\\
                                	&	11Mbps	&		2073.35 						&	[2059.62 ; 2087.08]\\
                                	&	Total		&		4149.03							&	[4139.91 ; 4158.15]\\
\cline{3-4}						      			&	Index		&\multicolumn{2}{c|}{0.9593866}\\
\hline
\multirow{3}{1.2cm}{PAS} 					& 5.5Mbps	&		1741.43							& [1733.81 ; 1749.05]\\
																	&	11Mbps	&		2782.73 						&	[2769.18 ; 2796.27]\\
																	&	Total		&		4524.16							&	[4514.01 ; 4534.31]\\
\cline{3-4}						      			&	Index		&\multicolumn{2}{c|}{0.9993147}\\
\hline
\end{tabular}
\caption{PAS with packet sizes uniformly distributed\label{tab:diff}}
\end{table}  

\subsection{Comparison with some other solutions} 

We have also compared PAS, our proposal, to other solution. The
results we obtained are presented hereafter.

\subsubsection{A simple backoff-based approach}

We have developed a simple backoff-based approach to solve the
performance anomaly. This approach is based on the solution proposed
by Heusse \emph{et Al.} \cite{Heu05}. 
The size of the contention window (CW) is adapted in the following
way: $CW = CW*\frac{11e6}{dataRate}$. 
In the simulations, the size of packets is uniformly distributed 
in the interval $[550;1450]$ bytes and there are
two emitters, one at transmitting at 5.5Mbps and the other at
11Mbps. 
Table~\ref{tab:CW} gives the average throughput as the average
fairness index. 
One can see that this approach is efficient, but not as 
efficient as our solution (see results for PAS in
Table~\ref{tab:diff}). 
This is due to the overhead introduced for each packet by the
backoff algorithm. 
Another problem of this approach is when small packets are sent 
by the fast station. In this case, the performance of the 
backoff-based approach decreases.



\begin{table}[!bp]
\centering
\begin{tabular}{|l|c|c|c|}
\hline
																	& 				& Th. (kbps) & Conf. Int. (0.05)\\
\hline
												 		& 5.5Mbps	&		1327.62							& [1314.12 ; 1341.11]\\
Backoff		             			&	11Mbps	&		3061.40 						&	[3045.48 ; 3077.32]\\
adaptation                	&	Total		&		4389.02							&	[4381.08 ; 4396.96]\\
\cline{3-4}						      &	Index		&\multicolumn{2}{c|}{0.9590798}\\
\hline
\end{tabular}
\caption{Backoff-based approach\label{tab:CW}}
\end{table}  

\subsubsection{Packet Division approach}

We have also tested the packet division approach proposed by Iannone
\emph{et Al.} \cite{IF05}. 
The simulations are carried out with two emitters, one transmitting at
11Mbps and the other at 5.5Mbps. 
The packet size of the fast station is set to 1500 bytes, while
the packet size of the slow station is but set to 727 bytes due to 
the fragmentation required in this solution. 
In the simulation, the two packet sizes are set to
1500 bytes with PAS. 
Table~\ref{tab:div} shows the results of these simulations. 
One can see from this table that the packet division
approach is less efficient, due to the overhead introduced by the
backoff and the header. 
It would also be trivial to show that when all wireless stations in 
the network use a small data rate, the network performance is reduced
because the packet fragmentation increases the payload/header ratio.  

\begin{table}[!tbp]
\centering
\begin{tabular}{|l|c|c|c|}
\hline
																	& 				& Th. (kbps) & Conf. Int. (0.05)\\
\hline
Packet										& 5.5Mbps	&		1779.97							& [1771.88 ; 1788.06]\\
division               		&	11Mbps	&		2377.61 						&	[2365.28 ; 2389.94]\\
					             		&	Total		&		4157.59							&	[4149.42 ; 4165.75]\\
\cline{3-4}						      &	Index		&\multicolumn{2}{c|}{0.9960047}\\
\hline
\multirow{3}{*}{PAS} 			& 5.5Mbps	&		1772.22							& [1764.16 ; 1780.29]\\
                          &	11Mbps	&		2936.01 						&	[2922.13 ; 2949.89]\\
                          &	Total		&		4708.24							&	[4698.97 ; 4717.51]\\
\cline{3-4}					      &	Index		&\multicolumn{2}{c|}{0.9980492}\\
\hline
\end{tabular}
\caption{Packet division approach\label{tab:div}}
\end{table}  


\subsubsection{Fixed time aggregation  approach} 

To carry out this simulation we have modified our implementation of
PAS, introducing a fixed $t\_p\_max=8000\mu s$. 
With this value, a node transmitting a 1500bytes data at 1Mbps can 
send only one packet. 
One can see from Table~\ref{tab:fixe}, comparing to 
Table~\ref{tab:diff}, that the aggregation using fixed time is more
efficient than our approach. 
This is due to the fact that, differently from PAS, the aggregation 
is always used.
On the other hand, this permanent aggregation implies longer delays 
between bursts. 
Table~\ref{tab:fix_burst} shows the number of bursts and the average time
between two bursts emitted by the same station. One can see from this
table that the delay induced by PAS is much smaller compared to the
other approach.   


\begin{table}[!tbp] 
\centering
\begin{tabular}{|l|c|c|c|}
\hline
																	& 				& Th. (kbps) & Conf. Int. (0.05)\\
\hline
\multirow{3}{*}{FIXED} 						& 5.5Mbps	&		1972.00							& [1955.38 ; 1988.62]\\
                                	&	11Mbps	&		2988.83 						&	[2959.72 ; 3017.94]\\
                                	&	Total		&		4960.84							&	[4947.75 ; 4973.92]\\
\cline{3-4}					      				&	Index		&\multicolumn{2}{c|}{0.9999999}\\
\hline
\end{tabular}
\caption{Fixed aggregation time\label{tab:fixe}}
\end{table}

\begin{table}[!bp]
\centering
\begin{tabular}{|l|c|c|c|}
\hline
																	& 				& Nb burts & Avg inter-access\\
\hline
\multirow{3}{*}{FIXED} 						& 5.5Mbps	&		7123							& 11230.07 $\mu$s\\
                                	&	11Mbps	&		6666 							&	12000.80 $\mu$s\\
\hline
\multirow{3}{*}{PAS} 							& 5.5Mbps	&		19570							& 4087.80 $\mu$s\\
                                	&	11Mbps	&		19346 						&	4135.11 $\mu$s\\
\hline
\end{tabular}
\caption{Performance anomaly delay results\label{tab:fix_burst}}
\end{table}  


\subsection{PAS in a multi-hop context}

\subsubsection{3 pairs scenario}
\begin{figure}[!tpb]
\centering
\includegraphics[scale=0.5]{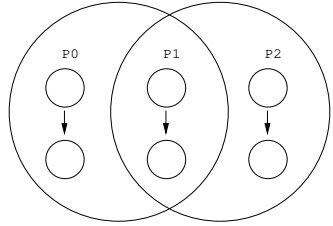}
\caption{The 3 pairs scenario\label{fig:3pairs}}
\end{figure}

\begin{figure}[tpb]
\centering
\includegraphics[scale=0.5]{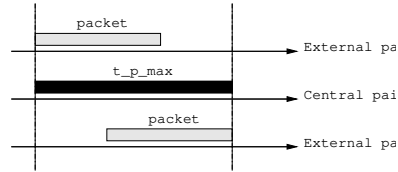}
\caption{The medium occupancy perceived by the central pair\label{fig:3pairs_occ}}
\end{figure}

Since all the mechanisms in PAS are fully distributed, PAS can also work in
a multi-hop context, where the wireless stations do not perceive the
same medium occupancy. 
If we consider the scenario depicted in Figure~\ref{fig:3pairs} we can 
see that the external pairs are fully independent. 
In this scenario, the central pair accesses the medium
95\% less than the external pairs, as demonstrated by Chaudet \emph{et
Al.} \cite{PE-WASUN04:Isa}. 
The medium occupancy perceived by the central pair is given in
Figure~\ref{fig:3pairs_occ}. 
One can see from this figure that the value of $t\_p\_max$ for the 
central pair can be at most equal to $t\_p1+t\_p2$, 
where $t\_pi_{i \in \{1,2\}}$ is the time needed for the
pair $i$ to transmit its packet. 
It is important to remark that here the maximum medium
occupancy time does not specifically correspond to a packet
transmission time. 
Table~\ref{tab:3pairs} shows the results on the 3 pairs scenario 
where the external pairs send 1000 bytes of data at 2Mbps and the 
central pair sends 1000 bytes of data at 11Mbps.

\begin{table}[!bp] 
\centering
\begin{tabular}{|l|c|c|c|}
\hline
																	& 				& Th. (kbps) 					& Conf. Int.		 \\
\hline
\multirow{3}{*}{PAS} 							& P0	&		1592.49 & [ 1584.16 ;  1600.82]\\
																	&	P1	&		102.21 & [   68.28 ;   136.15]\\
																	&	P2	&		1592.49 & [ 1584.09 ;  1600.89]\\
\hline
\multirow{3}{*}{802.11}						& P0	&	1634.15 & [ 1632.03 ;  1636.27]	\\
																	&	P1	& 6.44 & [    1.78 ;    11.11]		\\
																	&	P2	&	1632.86 & [ 1630.23 ;  1635.49]	\\
\hline
\end{tabular}
\caption{Results on 3 pairs scenario\label{tab:3pairs}}
\end{table}  

One can see from this table that even if PAS does not solve the
problem, the throughput of the central pair is highly
improved. 
Nevertheless, in this scenario a temporal fairness can not 
solve the problem and it seems necessary to modify the 802.11 
medium access control in order to provide each node the same 
probability to access the medium.

\subsubsection{Hidden terminals}

In Section~\ref{sec:des}, we have proposed a RTS/CTS mechanism for
PAS. Table~\ref{tab:rts_valid} evaluates this mechanism. In this
simulation we simulate two hidden nodes.
The RTS/CTS threshold is set to 200 bytes
and packet size to 1000 bytes. One can see from this table that the
RTS/CTS mechanism of PAS is close to the original 802.11's one.


\begin{table}[tbp] 
\centering
\begin{tabular}{|l|c|c|c|}
\hline
																	& 				& Th. (kbps) & Conf. Int. (0.05)\\
\hline
											 						& 11Mbps	&	1821.80 & [1770.05 ; 1873.55]	\\
802.11                           	&	11Mbps	&	1756.10 & [1704.39 ; 1807.82]	\\
RTS/CTS                          	&	Total		&	3577.91 & [3572.61 ; 3583.20]	\\
\cline{3-4}					      				&	Index		&\multicolumn{2}{c|}{0.9996629}\\
\hline
																	& 11Mbps	&	1760.83 & [1704.99 ; 1816.67]	\\
PAS                              	&	11Mbps	&	1818.07 & [1761.90 ; 1874.23]	\\
RTS/CTS                          	&	Total		&	3578.90 & [3573.59 ; 3584.21]	\\
\cline{3-4}					      				&	Index		&\multicolumn{2}{c|}{0.9997443}\\
\hline
\end{tabular}
\caption{RTS/CTS validation\label{tab:rts_valid}}
\end{table} 

In order to evaluate the performance of PAS in a multi-hop context
with aggregation, one of the hidden nodes uses a data rate of $x$,
where $x \in \{1,2,5.5\}$Mbps, while the other sends at $11$Mbps. 
Tables~\ref{tab:rts_55}, \ref{tab:rts_2} and \ref{tab:rts_1}
show the simulation results from these simulations. We can see from
these tables that PAS is more efficient and fairer than 802.11 when
one of the pairs has a data rate of $1$ or $2$Mbps. This is because
more aggregated packet can be sent by the fast station.  On the other,
we see that the results of PAS at $11$ and $5.5$Mbps are very close to
the ones of 802.11 (Table~\ref{tab:rts_55}). Since the time duration
in the RTS corresponds to the transmission time of the packet to send,
then a collision is likely to occur on the second packet of the
aggregated series. With $11$ and $5.5$Mbps, $t\_my\_left$ is not large
enough to aggregate the packet again, whereas with $11$ and $2$Mbps
(Table~\ref{tab:rts_2}) or $11$ and $1$Mbps (Table~\ref{tab:rts_1}),
$t\_my\_left$ is large enough to aggregate the packet that has
collided. In these two latter configurations, after some collisions,
the contention window of the slow station is large enough to allow the
aggregated sending of the fast station. 
 
\begin{table}[tbp] 
\centering
\begin{tabular}{|l|c|c|c|}
\hline
																	& 				& Th. (kbps) & Conf. Int. (0.05)\\
\hline
											 						& 5.5Mbps	&	1558.51 & [1518.96 ; 1598.06]	\\
802.11                           	&	11Mbps	&	1503.17 & [1450.54 ; 1555.80]	\\
RTS/CTS                          	&	Total		&	3061.68 & [3048.00 ; 3075.36]	\\
\cline{3-4}					      				&	Index		&\multicolumn{2}{c|}{0.9795797}\\
\hline
																	& 5.5Mbps	&	1584.43 & [1539.41 ; 1629.44]	\\
PAS                              	&	11Mbps	&	1463.86 & [1404.63 ; 1523.08]	\\
RTS/CTS                          	&	Total		&	3048.28 & [3033.50 ; 3063.06]	\\
\cline{3-4}					      				&	Index		&\multicolumn{2}{c|}{0.9733833}\\
\hline
\end{tabular}
\caption{RTS/CTS with 5.5 and 11Mbps nodes\label{tab:rts_55}}
\end{table} 

\begin{table}[bp] 
\centering
\begin{tabular}{|l|c|c|c|}
\hline
																	& 				& Th. (kbps) & Conf. Int. (0.05)\\
\hline
											 						& 2Mbps		&	1003.95 & [979.74 ; 1028.16]	\\
802.11                           	&	11Mbps	&	1064.32 & [1003.33 ; 1125.30]	\\
RTS/CTS                          	&	Total		&	2068.27 & [2031.39 ; 2105.14]	\\
\cline{3-4}					      				&	Index		&\multicolumn{2}{c|}{0.8721524}\\
\hline
																	& 2Mbps		&	827.97 	& [802.97 ;  852.97]	\\
PAS                              	&	11Mbps	&	1526.34 & [1463.53 ; 1589.15]	\\
RTS/CTS                          	&	Total		&	2354.31 & [2316.41 ; 2392.20]	\\
\cline{3-4}					      				&	Index		&\multicolumn{2}{c|}{0.9856836}\\
\hline
\end{tabular}
\caption{RTS/CTS with 2 and 11Mbps nodes\label{tab:rts_2}}
\end{table}

\begin{table}[tbp] 
\centering
\begin{tabular}{|l|c|c|c|}
\hline
																	& 				& Th. (kbps) & Conf. Int. (0.05)\\
\hline
											 						& 1Mbps		&	670.43 	& [658.39 ; 682.48]	\\
802.11                           	&	11Mbps	&	663.54 	& [611.37 ; 715.71]	\\
RTS/CTS                          	&	Total		&	1333.98 & [1293.81 ; 1374.15]	\\
\cline{3-4}					      				&	Index		&\multicolumn{2}{c|}{0.7205043}\\
\hline
																	& 1Mbps		&	552.45 & [535.59 ; 569.31]	\\
PAS                              	&	11Mbps	&	1237.35 & [1161.59 ; 1313.12]	\\
RTS/CTS                          	&	Total		&	1789.80 & [1730.87 ; 1848.73]	\\
\cline{3-4}					      				&	Index		&\multicolumn{2}{c|}{0.9071351}\\
\hline
\end{tabular}
\caption{RTS/CTS with 1 and 11Mbps nodes\label{tab:rts_1}}
\end{table}

Table~\ref{tab:rts_diff} shows the simulation results for two hidden
nodes transmitting at $1$ and $11$Mbps, with a packet size 
uniformly distributed between $[550;1450]$ bytes. 
In this simulation we set the RTS threshold to 1000 bytes. 
One can see from these results that, even with
different packet sizes, thus with a different RTS/CTS policy for
each packet (the RTS/CTS is not always activated), PAS is more
efficient and fair than 802.11. 
Note that in this simulation, the value of $t\_p\_max$ when 
RTS/CTS are not used corresponds to the transmission time of 
the acknowledgment.

\begin{table}[tbp] 
\centering
\begin{tabular}{|l|c|c|c|}
\hline
																	& 				& Th. (kbps) & Conf. Int. (0.05)\\
\hline
											 						& 1Mbps		&	305.92 & [297.37 ; 314.46]\\
802.11                           	&	11Mbps	&	919.55 & [888.74 ; 950.36]\\
RTS/CTS                          	&	Total		&	1225.47 & [1200.47 ; 1250.47]\\
\cline{3-4}					      				&	Index		&\multicolumn{2}{c|}{0.7945568}\\
\hline
																	& 1Mbps		&	226.39 & [218.78 ; 234.00]\\
PAS                              	&	11Mbps	&	1304.61 & [1269.65 ; 1339.57]\\
RTS/CTS                          	&	Total		&	1531.01 & [1500.21 ; 1561.80]\\
\cline{3-4}					      				&	Index		&\multicolumn{2}{c|}{0.9493314}\\
\hline
\end{tabular}
\caption{RTS/CTS with 1 and 11Mbps nodes with uniformly distributed
packet and 1000 bytes threshold\label{tab:rts_diff}}
\end{table}




%% file: conclusion.tex
In this paper we propose PAS, a dynamic packet aggregation mechanism 
to solve the performance anomaly of 802.11. 
Our solution is based on the fact that the same transmission time is 
given to each station. 
This transmission time is computed dynamically and is equal to the
maximum occupation time perceived on the medium.  
When a node has the opportunity to use the channel, it sends as many 
packets as the transmission time allows. 
The aggregation is done by waiting only for a SIFS period 
between the reception of an ACK and the beginning of the next
transmission. 
To increase the dynamicity and to reduce the convergence
time, the transmission time is set to $0$ after each successful 
transmission (or burst of aggregate transmission).

We have shown, through both analytical analysis and  simulation, 
that our protocol solves the performance anomaly in many scenarios. 
The aggregate throughput can be increased and the 
time-based fairness is almost reached in almost every of the tested 
configurations. 
We have also shown that our approach does not need extra information
than that already furnished by IEEE 802.11 standard, thus it 
can be easily implemented. 
An important characteristic of our proposal is the fact that
it can be also used in multi-hop networks, improving also in this
scenario the performances.